\documentclass[prl,twocolumn,amsmath,nobibnotes,nofootinbib,superscriptaddress,preprintnumbers,hyperref]{revtex4-1}
\usepackage{graphicx}
\usepackage{multirow} 
\usepackage[usenames,dvipsnames]{xcolor}
\usepackage[breaklinks, colorlinks, citecolor=blue, linkcolor=blue, urlcolor=blue]{hyperref}
\usepackage{txfonts}

\newcommand{\LCDM}{\Lambda\text{CDM}}
\newcommand{\Treg}{\mathrm{T_{reg}}}
\newcommand{\Tirr}{\mathrm{T_{irr}}}
\newcommand{\OmL}{\Omega_{\Lambda}}
\newcommand{\sigmaH}{\ensuremath{(\sigma/H)_0}}
\newcommand{\sigmaSH}{\ensuremath{(\sigma_S/H)_0}}
\newcommand{\sigmaVH}{\ensuremath{(\sigma_V/H)_0}}
\newcommand{\sigmaTregH}{\ensuremath{(\sigma_{T,\rm reg}/H)_0}}
\newcommand{\sigmaTirrH}{\ensuremath{(\sigma_{T,\rm irr}/H)_0}}
\newcommand{\mK}{\mathrm{mK}}
\begin{document}

\title{How isotropic is the Universe?}

\author{Daniela Saadeh}
\email{daniela.saadeh.13@ucl.ac.uk}
\affiliation{Department of Physics and Astronomy, University College London, London WC1E 6BT, U.K.}

\author{Stephen M. Feeney}
\affiliation{Astrophysics Group, Imperial College London, Blackett Laboratory, Prince Consort Road, London, SW7 2AZ, U.K.}

\author{Andrew Pontzen}
\affiliation{Department of Physics and Astronomy, University College London, London WC1E 6BT, U.K.}
\author{Hiranya V. Peiris}
\affiliation{Department of Physics and Astronomy, University College London, London WC1E 6BT, U.K.}

\author{Jason D. McEwen}
\affiliation{Mullard Space Science Laboratory (MSSL), University College London, Surrey RH5 6NT, U.K.}

\date{\today}

\begin{abstract}
A fundamental assumption in the standard model of cosmology is that the Universe is isotropic on large scales. Breaking this assumption leads to a set of solutions to Einstein's field equations, known as Bianchi cosmologies, only a subset of which have ever been tested against data. For the first time, we consider all degrees of freedom in these solutions to conduct a general test of isotropy using cosmic microwave background temperature and polarization data from \textit{Planck}. For the vector mode (associated with vorticity), we obtain a limit on the anisotropic expansion of $\sigmaVH < 4.7 \times 10^{-11}$ (95\% CI), which is an order of magnitude tighter than previous \textit{Planck} results that used CMB temperature only. We also place upper limits on other modes of anisotropic expansion, with the weakest limit arising from the regular tensor mode, $\sigmaTregH<1.0 \times 10^{-6}$ (95\% CI). Including all degrees of freedom simultaneously for the first time, anisotropic expansion of the Universe is strongly disfavoured, with odds of 121,000:1 against.
\end{abstract}

%\keywords{}

\maketitle

The standard $\Lambda$CDM model of cosmology assumes the Copernican principle, which states that the Universe is isotropic  and homogeneous on large scales.
In this work, we test whether the expansion of the universe is indeed isotropic, using cosmic microwave background (CMB) data from the \textit{Planck} satellite. 

Assuming homogeneity and isotropy, the solutions to Einstein's field equations are given by the Friedmann-Lema\^{i}tre-Robertson-Walker (FLRW) metric. Relaxing the isotropy requirement while continuing to demand homogeneity leads instead to Bianchi metrics \cite{Bianchi_originale,Pontzen:2016}. The anisotropic expansion in these models imprints a signal in the CMB since photons redshift at different rates depending on their direction of travel \cite{Hawking69,Barrow}, an effect known as {\it shear}. The CMB can therefore be used to place limits on anisotropic expansion, although to do so the geometric signal must be disentangled from the stochastic fluctuations 
responsible for structure formation.

Before the temperature fluctuations of the CMB had been characterized, it was possible to place preliminary upper limits on the magnitude of anisotropy \cite{Doroshkevich73,CollinsHawking73}.
Later tests for anisotropic expansion (see, e.g., Refs.~\cite{bunn:1996, Kogut97, Jaffe_et_al_2005, Bridges2007, Jason, Planck_background_2013, Planck_background_2015}) focussed on vorticity (i.e., universal rotation) and thus tested only some of the ways in which the Universe can be anisotropic.  In this \textit{Letter}, we carry out the first general test using all shear degrees of freedom and the widest possible range of geometric configurations that describe anisotropy. We incorporate polarization data (as well as temperature) in the statistical analysis for the first time. This enables us to obtain order-of-magnitude tighter constraints on vorticity than previously obtained using \textit{Planck} data. The large number of physical and nuisance parameters necessary for a full exploration requires us to develop a new sampling package, \texttt{ANICOSMO}$_2$, based on \texttt{PolyChord} \cite{Polychord_1,Polychord_2}. Together, these developments allow us to perform the first general test of isotropic expansion by constraining the full set of Bianchi degrees of freedom with CMB data.

\textbf{Anisotropic models:} Departures from isotropy that preserve homogeneity are described by Bianchi models, which can be subdivided into a number of ``types'' describing the overall geometry of space.
One may show \cite{EllisMaccallum69} that only certain types allow for an isotropic limit (specifically, types I and VII$_0$, V and VII$_h$, and IX contain flat, open and closed FLRW universes, respectively).
Among these, all but the closed models can be obtained from limits of the Bianchi VII$_h$ case \cite{Barrow}. The closed case induces only a quadrupole in the CMB temperature and polarization and, consequently, is difficult to constrain. In this work, we therefore consider the most general Bianchi VII$_h$ freedom --- including its sub-types VII$_0$, V and I --- allowing us to test for the most general departure from isotropy that retains homogeneity within a flat or open Universe. 

In all Bianchi types, anisotropy is quantified in terms of the shear tensor $\sigma_{\mu\nu}$, which describes the deformation that a fluid element in the Universe undergoes as a result of anisotropic expansion. For small deviations from isotropy, the full shear freedom can be expressed as a set of five non-interacting modes that behave like scalars (S), vectors (V) or tensors (T) under rotations around a preferred axis of the Bianchi model~\cite{Lukash1976,P&C11}. Only vector modes, which generate vorticity, have previously been confronted with \textit{Planck} data~\cite{Planck_background_2013, Planck_background_2015}.

We model the energy content of the Universe as the sum of perfect fluids corresponding to matter, radiation and dark energy.  The Einstein equations then dictate the evolution of the scalar, vector and tensor shear components. We consistently include the fluid motion relative to the comoving frame\footnote{We do assume, however, that all sources are perfect fluids: this will be accurate despite anisotropic stresses in the radiation component since the calculations start long after matter-radiation equality.}.
 Scalar and vector modes decay steeply ($\propto a^{-3}$, where $a$ is the direction-averaged scale factor) as the Universe expands, whereas tensor modes can be characterized as the linear combination of modes that initially grow or decay, labelled `regular' ($\Treg$) or `irregular' ($\Tirr$) following Ref.~\cite{Pontzen09}. This initial behavior leads in both cases to an oscillatory solution, with a phase difference between the two modes. For a given shear amplitude today, initially-decaying modes are larger at recombination than initially-growing modes, and therefore imprint greater polarization anisotropy \cite{Rees68,MatznerTolman82,P&C07}; furthermore, in all but the scalar modes, $E$-mode polarization is efficiently converted into similar levels of $B$-mode polarization \cite{P&C07}. As a consequence, CMB polarization data are the ideal probe to constrain all but the regular tensor modes~\cite{method_paper}, and are expected to give rise to even stronger limits than temperature anisotropy or nucleosynthesis constraints~\cite{Pontzen:2016}.

\begin{figure}
 \includegraphics[width=\columnwidth]{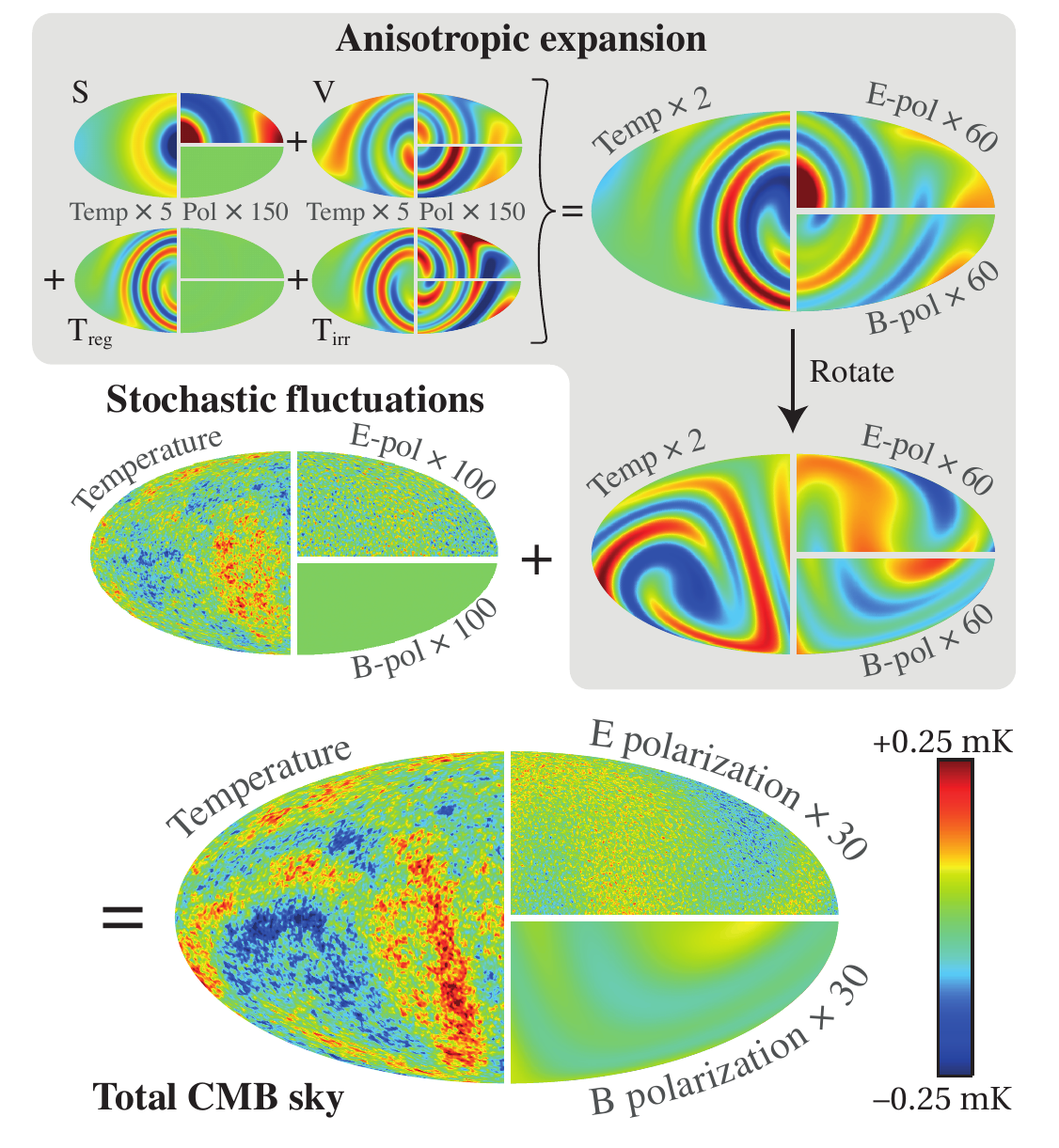}
 \caption{The CMB sky in the near-isotropic limit is formed from the addition of a standard, stochastic background for the inhomogeneities to a pattern arising from small anisotropic expansion. In this work, for the first time, we constrain all modes of the anisotropic expansion (scalar, vector, regular tensor, irregular tensor). Here we have depicted anisotropic expansion that is large compared to our limits (though still small compared to the isotropic mean) for illustrative purposes; specifically,  $\sigmaSH = 4.2 \times 10^{-10}$, $\sigmaVH = 3.2 \times 10^{-10}$, $\sigmaTregH = 1.1 \times 10^{-6}$, $\sigmaTirrH = 1.8 \times 10^{-8}$, with Bianchi scale parameter $x=0.62$. Each map shows temperature (left), $E$-mode polarization (upper right) and $B$-mode polarization (lower right). The overall temperature color scale for the bottom, final map is $-0.25\,\mK<T<0.25\,\mK$, with polarization amplitudes exaggerated by a factor 30 relative to this. Other panels have been rescaled as indicated, for clarity.}\label{fig:bianchi-plus-lcdm}
\end{figure}

Figure~\ref{fig:bianchi-plus-lcdm} summarizes the origin of CMB fluctuations. In the limit that the deviation from isotropy is small, the geometric and stochastic fluctuations can be added linearly. To compute the signal imprinted by the background anisotropy (shaded region of Fig.~\ref{fig:bianchi-plus-lcdm}), we have developed the Boltzmann-hierarchy code \texttt{ABSolve} \cite{method_paper}. \texttt{ABSolve} can predict temperature and polarization maps and power spectra for all the shear modes in Bianchi I, V, VII$_0$ and VII$_h$ and is designed to accurately characterize the deterministic Bianchi pattern across the whole parameter space considered \cite{method_paper}. To naturally allow types I, V and VII$_0$ within our parameter space we allow the rotation scale of the shear principal axes relative to the present-day horizon scale (denoted $x$ by convention) to become sufficiently large. Strictly, type V is obtained as $x \to \infty$; to accommodate this possibility in our prior space we found that $x_{\rm max}=10^5$ is sufficient for convergence. Similarly, the flat Bianchi VII$_0$ limit is obtained by allowing $\Omega_K \to 0$; we consider values down to $\Omega_{K,\rm min}=10^{-5}$. Bianchi type I is obtained as the $x$ and $\Omega_K$ limits are approached simultaneously.

\begin{table}
\begin{center}
\caption{Parameter priors. The first seven parameters are the $\LCDM$ baryon ($\Omega_bh^2$) and cold dark matter ($\Omega_ch^2$) physical densities, dark energy ($\Omega_{\Lambda}$) and curvature ($\Omega_K$) densities, scalar spectral index ($n_{\rm s}$) and amplitude ($A_{\rm s}$) and optical depth to reionization ($\tau$). The following ten parameters are Bianchi degrees of freedom: the rotation scale of the shear principal axes ($x$); the normalized shear scalar today $\sigmaH$ for scalar, vector, regular tensor and irregular tensor modes; the vector-to-tensor angular offset ($\gamma_{VT}$); three Euler angles defining the principal axis orientation ($\{\alpha,\beta,\gamma\}$) and the pattern's handedness ($p$). The remaining parameters ($y_{\rm cal}$, \textit{Planck}'s absolute map calibration, and $\Theta_{\rm high}$, a list of 14 parameters describing foreground and instrumental contaminants) are nuisance parameters used by the low- and high-$\ell$ likelihood functions, respectively. \label{Tab:priors}} 

 \begin{tabular}{cccc}
\hline
Parameter & Prior Range & Prior Type & Speed\tabularnewline
\hline 
$\Omega_bh^2$ & $[0.005,0.05]$ & uniform & 1 \tabularnewline
$\Omega_ch^2$ & $[0.05,0.3]$ & uniform & 1 \tabularnewline
$\OmL$ & $[0,0.99]$ & uniform & 1 \tabularnewline
$\Omega_K$ & $[10^{-5},0.5]$ & uniform & 1 \tabularnewline
$n_{\mathrm{s}}$ & $[0.9,1.05]$ & uniform & 1 \tabularnewline
$A_{\mathrm{s}}$ & $[1,5]\times10^{-9}$ & log-uniform & 1 \tabularnewline
$\tau$ & $[0.01,0.2]$ & uniform & 1 \tabularnewline
\hline
$x$ (vector-only search) & $[0.05,2]$ & uniform & 2 \tabularnewline
$x$ (all-mode search) & $[0.05,10^5]$ & log-uniform & 2 \tabularnewline
$\sigmaSH$ & $[-10^{-8},10^{-8}]$ & uniform & 2 \tabularnewline
$\sigmaVH$ & $[10^{-12},10^{-8}]$ & log-uniform & 2 \tabularnewline
$\sigmaTregH$ & $[10^{-12}$,$10^{-4}]$ & log-uniform  & 2 \tabularnewline
$\sigmaTirrH$ & $[10^{-12}$,$10^{-4}]$ & log-uniform & 2 \tabularnewline
$\gamma_{VT}$ & $[0^{\circ},180^{\circ}]$ & uniform & 2 \tabularnewline
$\alpha$ & $[0^{\circ},360^{\circ}]$ & uniform & 2 \tabularnewline
$\beta$ & $[0^{\circ},180^{\circ}]$ & sine-uniform & 2 \tabularnewline
$\gamma$ & $[0^{\circ},360^{\circ}]$ & uniform  & 2 \tabularnewline
$p$ & left/right & discrete  & N/A \tabularnewline
\hline
$y_{\rm cal}$ & \multicolumn{2}{c}{see Ref.~\cite{Planck_likelihood}} & 3 \tabularnewline
$\Theta_{\rm high}$ & \multicolumn{2}{c}{see Ref.~\cite{Planck_likelihood}} & 4 \tabularnewline
\hline
\end{tabular}
\end{center}
\end{table}

\textbf{Data: } To confront the model for anisotropic expansion described above with data, we redeveloped the \texttt{ANICOSMO} package~\cite{Jason}; our remodeled code, \texttt{ANICOSMO$_2$}, calculates the CMB contributions from anisotropic expansion using \texttt{ABSolve} (described above) and from inhomogeneities using \texttt{CAMB}~\cite{CAMB}. The new statistical approach combines a custom low-$\ell$ likelihood based on the \textit{Planck} $T$+$P$ low-$\ell$ likelihood with the standard \textit{Planck} temperature-only high-$\ell$ likelihood~\citep{Planck_likelihood}. The high dimensionality of the resulting parameter space, alongside high computational costs for a full likelihood evaluation, made it necessary to redesign \texttt{ANICOSMO} around the slice-sampling nested sampler \texttt{PolyChord}~\citep{Polychord_1, Polychord_2}. We will now describe each of these developments briefly in turn; further detail is given in the {\it Supplemental Materials}.

The low-$\ell$ likelihood, providing constraints on large angular scales from temperature and polarization, is applied to multipoles in the range $2<\ell<29$.\footnote{2015 CMB spectra and likelihood code section 2.2.1: \url{https://wiki.cosmos.esa.int/planckpla2015/index.php/CMB_spectrum_\%26_Likelihood_Code\#Low-.E2.84.93_likelihoods}} It is based on foreground-cleaned maps, downgraded to {\tt HEALPix}~\cite{Healpix} resolution $N_{\text{side}}=16$ and masked using the LM93 mask~\cite{Planck_likelihood}. The temperature map is generated by the \texttt{Commander} component-separation algorithm operating on data from the \textit{Planck} 30--857 GHz channels~\cite{Planck_maps}, nine-year WMAP 23--94 GHz channels~\cite{WMAP_likelihood} and 408 MHz observations \cite{408MHz_observations}. The polarization data are derived from \textit{Planck}'s 70 GHz maps, cleaned using its 30 GHz and 353 GHz channels as templates for low- and high-frequency contamination. Note that this represents only a small fraction of \textit{Planck}'s large-scale polarization data: the constraining power of \textit{Planck} data will increase with future releases. 

We modified the low-$\ell$ code described in Ref.~\citep{Planck_likelihood} to accept maps of the Bianchi temperature and polarization anisotropies as inputs. These maps, which include the \textit{Planck} beam, are computed (as described above) by \texttt{ABSolve}, then masked and concatenated into a single vector retaining only the unmasked $T$, $Q$ and $U$ pixels, where $T$, $Q$ and $U$ are Stokes parameters describing the CMB intensity and linear polarization.
The vector is divided by the \textit{Planck} map calibration $y_{\mathrm{cal}}$ since the absolute normalization is uncertain \cite{Planck_likelihood} (similarly, the \texttt{CAMB}-computed power spectra required by the low-$\ell$ likelihood are divided by $y_{\mathrm{cal}}^2$). 
Our final Bianchi vector is subtracted from the vector of \textit{Planck} data to correct for the anisotropic expansion corresponding to the input parameters.  

A direct calculation of the likelihood from the resulting corrected data vector is computationally prohibitive, even at modest resolution, due to the inversion of a large pixel covariance matrix that changes in response to the cosmological and calibration parameters. 
The original \textit{Planck} likelihood code optimizes the inversion for the case that the data vector does not change between evaluations. However, the anisotropic-expansion correction to the maps is parameter-dependent. We have therefore generalized the code to retain similar efficiency when all inputs are changing. For more information, see {\it Supplemental Materials}.

We employ the \textit{Planck} $TT$ high-$\ell$ power-spectrum likelihood~\citep{Planck_likelihood} for multipoles $29 < \ell \leq 2508$.\footnote{2015 CMB spectra and likelihood code section 2.2.2: \url{http://wiki.cosmos.esa.int/planckpla2015/index.php/CMB_spectrum_\%26_Likelihood_Code\#High-.E2.84.93_likelihoods}} This uses temperature data from various combinations of the 100--217 GHz detectors, masked to remove the Galactic plane, regions of high CO emission and point sources. The remaining astrophysical foregrounds are modeled within the \textit{Planck} code using 14 parameters. To take into account the imprint of anisotropic expansion on small scales, we sum the {\tt ABSolve} and {\tt CAMB} power spectra within \texttt{ANICOSMO}$_2$ before passing to the high-$\ell$ likelihood. For anisotropic models the power spectrum does not provide lossless data compression, but in the expected limit that the geometric signal is subdominant to the stochastic component, one may show that the approach gives a good approximation\footnote{Even for anisotropic theories one may estimate the full-sky power accurately from cut-sky data~\cite{PP10}.} to the correct likelihood (see Ref.~\cite{method_paper}).

In total, there are 32 parameters describing the cosmology, calibration and foregrounds (Table~\ref{Tab:priors}). To sample this high-dimensional space efficiently, we have redesigned our approach around the \texttt{PolyChord} package \cite{Polychord_1,Polychord_2}, which substitutes slice sampling for the rejection sampling~\cite{MultiNest1,MultiNest2,MultiNest3} employed in our previous work which sampled a maximum of 13 parameters \cite{method_paper,Jason}. Rejection sampling scales exponentially with dimensionality, whereas \texttt{PolyChord} scales at worst as $\mathcal{O}(D^3)$ with the further advantage of an algorithm which parallelizes nearly linearly.
\texttt{PolyChord} is also capable of exploiting likelihood optimizations arising from fixing some parameters. Recalculating the {\tt CAMB} power spectra, {\tt ABSolve} maps and spectra, low-$\ell$ likelihood, and high-$\ell$ likelihood take approximately 40, 3, 0.5 and 0.006 seconds respectively, on a single core. \texttt{PolyChord} efficiently explores the parameter space by oversampling the faster foreground parameters with respect to the slower cosmological parameters.\footnote{We set up \texttt{PolyChord} such that $n_{\rm live}=500$, $n_{\rm repeats}=8$, and it spends 70, 20, 5, and 5\% of its time sampling parameters of each speed (see Table~\ref{Tab:priors}).}
We marginalize over the handedness, $p$, of the Bianchi models by sampling the left- and right-handed posteriors individually and combining the results as described in the {\it Supplemental Materials}. The priors applied have been motivated in Ref.~\cite{method_paper}.

\textbf{Results: } As a test of the information carried by the polarization, we first apply our search to vector modes only, 
as vorticity has been the focus of all previous work constraining anisotropic expansion with \textit{Planck} \cite{Planck_background_2015,Planck_background_2013}. We obtain a limit of $\sigmaVH < 4.9 \times10^{-11}$ (95\% CI). 
This can be recast in terms of the vorticity parameter $(\omega/H)_0$, which expresses the rotation rate of the universe, giving $(\omega/H)_0 < 5.2 \times 10^{-11}$ (95\% CI). Although constraints are slightly prior-dependent (see Ref. \cite{method_paper} for a discussion), previous analyses \cite{Planck_background_2015,Planck_background_2013} report limits on the vorticity of $(\omega/H)_0<7.6 \times 10^{-10}$ (95\% CI). Our new limit is therefore tightened relative to earlier constraints by an order of magnitude.

\begin{table*}
\begin{center}
\caption{95\% credible intervals for the anisotropy modes and log-evidence ratios for the overall anisotropic models compared to homogeneous and isotropic flat $\LCDM$.  Negative values of the log-evidence ratio favor isotropy.  \label{Tab:anisotropy_constraints}}
\begin{tabular}{ccc}
\hline
Mode & \textit{Planck} & WMAP \\
\hline
Scalar & $-6.7 \times 10^{-11} < \sigmaSH < 9.6 \times 10^{-11}$ & $-3.5 \times 10^{-10} < \sigmaSH < 4.0 \times 10^{-10}$ \\
Vector & $\sigmaVH < 4.7 \times 10^{-11}$ & $\sigmaVH < 1.7 \times 10^{-10}$ \\
Tensor, reg & $\sigmaTregH < 1.0 \times 10^{-6}$ & $\sigmaTregH < 1.3 \times 10^{-6}$ \\
Tensor, irreg & $\sigmaTirrH < 3.4 \times 10^{-10}$ & $\sigmaTirrH < 6.7 \times10^{-10}$ \\
\hline
Vector (vorticity) only & $-5.6 \pm 0.3$ & $-3.3 \pm 0.1$ \tabularnewline
All anisotropic modes & $-11.7 \pm 0.3$ & $-8.0 \pm 0.2$ \tabularnewline
\hline
\end{tabular}
\end{center}
\end{table*}

For the first full test of isotropy, our second analysis simultaneously constrains all degrees of freedom in the cosmological shear tensor. The resulting limits are presented in Table~\ref{Tab:anisotropy_constraints}. We also present the constraints calculated using our older likelihood \cite{method_paper,Jason}, based on temperature data from the {\it Wilkinson Microwave Anisotropy Probe} (WMAP) internal linear combination maps \cite{WMAP9yr2013}, as a baseline for comparison. Note that the WMAP setting in Ref.~\cite{method_paper} already contained some methodological improvements (specifically the treatment of high-$\ell$ Bianchi power) that enhanced the constraints over standard analyses. However, because we also widen the prior range on $x$ (Table \ref{Tab:priors}) to include the Type V limit, the all-mode constraints are not directly comparable to results from these older single-mode searches.

We recover upper limits for all modes, showing that the Universe is consistent with isotropic expansion. The tightest constraints are placed on the fastest-decaying modes: the scalar, vector and irregular tensor modes. The limits on the regular tensor modes are much less stringent as a result of the dynamics. For most modes, the shear at last scattering is amplified by a factor $a^{-3}$ relative to the present-day value; for the regular tensors, this enhancement factor can be vastly smaller. The temperature and, especially, polarization anisotropies corresponding to a fixed present-day shear are therefore also smaller. The effect on polarization is pronounced, making such data particularly effective at discriminating between the two tensor modes, for which the temperature pattern is generally similar.

The consistency of the data with statistical isotropy is also borne out by comparing the model-averaged likelihoods (known as evidences) for the Bianchi cosmology and flat $\LCDM$. The ratio of the model-averaged likelihoods tells us whether the Universe is most likely anisotropic or isotropic, given our CMB observations. The bottom two rows of Table \ref{Tab:anisotropy_constraints} contain the ratios calculated for our vector-only and all-modes analyses. 
Upgrading from WMAP temperature data to \textit{Planck} data with polarization, the preference against anisotropic expansion becomes significantly stronger, with odds of 270:1 against anisotropy in the vector-only case. In the all-modes analysis, the larger parameter space leads to overwhelming odds against anisotropic expansion: 121,000:1.

\textbf{Conclusions: } In this work, we put the assumption that the Universe expands isotropically to its most stringent test to-date. For the first time, we searched for signatures of the most general departure from isotropy that preserves homogeneity in an open or flat universe, without restricting to specific degrees of freedom. We have remodeled existing frameworks to analyze CMB polarization data in addition to temperature, allowing us to place the tightest constraints possible with the current CMB data. 
We find overwhelming evidence against deviations from isotropy, placing simultaneous upper limits on all modes for the first time, and improving \textit{Planck} constraints on vorticity by an order of magnitude.

\begin{acknowledgments}
DS thanks Franz Elsner, Boris Leistedt, Sabino Matarrese, Alessandro Renzi for useful discussions. SMF thanks Will Handley for assistance with the {\tt PolyChord} package.
DS is supported by the Perren Fund and the IMPACT fund and partially supported by the RAS Small Grant Scheme. SMF is supported by the Science and Technology Facilities Council in the UK. AP is supported by the Royal Society. HVP is partially supported by the European Research Council under the European Community's Seventh Framework Programme (FP7/2007-2013) / ERC grant agreement number 306478-CosmicDawn. JDM is partially supported by the Engineering and Physical Sciences Research Council (grant number EP/M011852/1).
We acknowledge use of the Legacy Archive for Microwave Background Data Analysis (LAMBDA). Support for LAMBDA is provided by the NASA Office of Space Science.
Based on observations obtained with \textit{Planck} (\url{http://www.esa.int/Planck}), an ESA science mission with instruments and contributions directly funded by ESA Member States, NASA, and Canada.
\end{acknowledgments}

\bibliography{Bianchi_letter}

\appendix

\section*{Supplemental Materials}

In this Section we further describe the modifications we have made to the \textit{Planck} likelihood for this study, and our method for marginalizing over the handedness of the Bianchi models. In our model, the observed CMB maps $\mathbf{m}$ (in this case, a vector containing only the unmasked $T$, $Q$ and $U$ pixels) can be decomposed into a deterministic Bianchi template $\mathbf{b}$, stochastic $\LCDM$ fluctuations $\mathbf{s}$ and Gaussian instrumental noise $\mathbf{n}$
\begin{equation}
 \mathbf{m}=\mathbf{b}(\Theta_{\rm B})+\mathbf{s}(\Theta_{\LCDM})+\mathbf{n},
\end{equation}
where $\Theta_{\rm B}$ and $\Theta_{\LCDM}$ are the Bianchi and $\LCDM$ parameters, respectively. The likelihood function then takes the form of a Gaussian with mean set by the Bianchi template and covariance matrix $\mathbf{M}$ defined by the stochastic fluctuations and instrumental noise properties:
\begin{equation}
P(\mathbf{m}|\Theta_{\rm B},\Theta_{\LCDM}) = \frac{\exp\left[ -\frac{1}{2}(\mathbf{m}-\mathbf{b})^{\rm T} \mathbf{M}^{-1} (\mathbf{m}-\mathbf{b}) \right]}{|2\pi\mathbf{M}|^{1/2}}.
\label{Eq:likelihood}
\end{equation}

Calculating this likelihood is computationally intensive, even at modest resolution, due to the inversion of the large pixel covariance matrix $\mathbf{M}$. The Planck Collaboration therefore decompose the covariance matrix in order to make the computation more convenient~\citep{Planck_likelihood}. The matrix $\mathbf{M}$ is first split (at $\ell_{\rm cut}$) into a varying low-$\ell$ cosmology-dependent matrix and a fixed high-$\ell$ correlated noise matrix $\mathbf{M}_0$. The low-$\ell$ matrix, whose rank ($N_\lambda = 3[(\ell_{\rm cut} + 1)^2 - 4]$) can be much smaller than that of the full pixel covariance ($N_{\rm pix}$), is then further decomposed to allow the use of the Woodbury identities
\begin{eqnarray}
&{\mathbf M}^{-1} = {\mathbf M}_0^{-1} - {\mathbf M}_0^{-1}{\mathbf V}^T\left({\mathbf A}^{-1}+{\mathbf V}{\mathbf M}_0^{-1}{\mathbf V}^T\right)^{-1}{\mathbf V}{\mathbf M}_0^{-1} \nonumber \\
&|{\mathbf M}| = |{\mathbf M}_0|| {\mathbf A}|| {\mathbf A}^{-1}+{\mathbf V}{\mathbf M}_0^{-1}{\mathbf V}^T|  \label{Eq:Woodbury_inverse}
\end{eqnarray}
\\
to rapidly recalculate the inverse and determinant of $\mathbf{M}$. Here, ${\mathbf A}(\Theta_{\LCDM})$ is a block-diagonal $N_\lambda \times N_\lambda$ matrix encoding the cosmology dependence, and ${\mathbf V}$ is an $N_\lambda \times N_{\rm pix}$ projection matrix. The unmodified ($\mathbf{b} = 0$) \textit{Planck} low-$\ell$ likelihood code precomputes ${\mathbf m}^T{\mathbf M}_0^{-1}{\mathbf m}$, ${\mathbf V}{\mathbf M}_0^{-1}{\mathbf m}$ and ${\mathbf V}{\mathbf M}_0^{-1}{\mathbf V}^T$ to save time, then discards the data, covariance and projection matrix. We have modified the likelihood code to retain these quantities, as they are needed to calculate the likelihood~\eqref{Eq:likelihood} using the decomposed covariance~\eqref{Eq:Woodbury_inverse} in the presence of a Bianchi component.

Even after speeding up the inverse covariance matrix computation, this pixel-based approach is only feasible at the largest scales; however, neglecting the small scales discards cosmological information that is highly constraining not only for the $\LCDM$ component, but also the anisotropic background~\cite{method_paper}. For this reason, we add high-$\ell$ information in the form of the \textit{Planck} $TT$ high-$\ell$ power-spectrum likelihood~\citep{Planck_likelihood} for $\ell_{\rm cut} < \ell \leq \ell_{\rm max}$. In this case, the only modification required by the presence of a Bianchi component is to pass as input the summed Bianchi and $\LCDM$ power spectra, calculated using \texttt{ABSolve} and \texttt{CAMB} respectively. For anisotropic models the power spectrum does not provide lossless data compression, but in the limit where the Bianchi signal is subdominant to $\LCDM$, this gives a good approximation to the exact likelihood in Eq.~\ref{Eq:likelihood} (see Ref.~\cite{method_paper}, Appendix A).

To present conclusions marginalized over the handedness of the Bianchi models, we sample the posterior for each handedness separately. Denoting the evidence for, e.g., left handedness as $E_{\rm left} = P({\mathbf m}|p={\rm left})$, Bayes' theorem implies that the evidence for a model allowing both handednesses is
\begin{equation}
E = (E_{\rm left} + E_{\rm right}) / 2,
\end{equation}
and the joint posterior on this model's parameters is
\begin{equation}
P(\Theta|\mathbf{m}) = \frac{ E_{\rm left} P(\Theta|\mathbf{m},p={\rm left}) + E_{\rm right} P(\Theta|\mathbf{m},p={\rm right})}{ E_{\rm left} + E_{\rm right} }.
\end{equation}
All limits and evidence values (Table \ref{Tab:anisotropy_constraints}) have been quoted as a joint posterior in this way.

\end{document}